\overfullrule=0pt
\input harvmac
\def\a{{\alpha}}

\def\l{{\lambda}}

\def\b{{\beta}}

\def\g{{\gamma}}

\def\d{{\delta}}
\def\e{{\epsilon}}
\def\s{{\sigma}}
\def\k{{\kappa}}

\def\half{{1\over 2}}
\def\p{{\partial}}

\def\t{{\theta}}

\Title{\vbox{\hbox{IFT-P.034/2005 }}}
{\vbox{
\centerline{\bf Equivalence of Two-Loop Superstring Amplitudes }
\centerline{\bf in the Pure Spinor and RNS Formalisms }}}
\bigskip\centerline{Nathan Berkovits\foot{e-mail: nberkovi@ift.unesp.br}
and Carlos R. Mafra\foot{e-mail: crmafra@ift.unesp.br}}
\bigskip
\centerline{\it Instituto de F\'\i sica Te\'orica, Universidade Estadual
Paulista}
\centerline{\it Rua Pamplona 145, 01405-900, S\~ao Paulo, SP, Brasil}

\vskip .3in
The pure spinor formalism for the superstring has recently been used to
compute massless four-point two-loop amplitudes in a manifestly
super-Poincar\'e covariant manner. In this paper, we show that when
all four external states are Neveu-Schwarz, the two-loop amplitude
coincides with the RNS result.

\vskip .3in

\Date {September 2005}

\newsec{Introduction}

String theory is currently the most promising model for
unification of the forces. In bosonic string theory, the prescription
for computing 
perturbative scattering amplitudes is well-developed and has been
used to compute amplitudes with arbitrary numbers of loops. Unfortunately,
these multiloop amplitudes suffer from unphysical divergences which make
bosonic string theory inconsistent. In superstring theory, spacetime
supersymmetry helps in cancelling these divergences. 
However, because
spacetime supersymmetry is not manifest in the 
Ramond-Neveu-Schwarz (RNS) formalism \ref\rns{P. Ramond,
{\it Dual Theory for Free Fermions}, Phys. Rev. D3 (1971) 2415\semi
A. Neveu and J.H. Schwarz, {\it Factorizable Dual Model of Pions},
Nucl. Phys. B31 (1971) 86.} for the superstring,
it is difficult to explicitly prove the cancellation of
divergences using this formalism.
Although the Green-Schwarz (GS) formalism \ref\gs
{M.B. Green and J.H. Schwarz, {\it Covariant Description of Superstrings},
Phys. Lett. B136 (1984) 367.} for the superstring is
manifestly spacetime supersymmetric, its non-quadratic action makes
it difficult to quantize except in light-cone gauge.

Five years ago, a new formalism for the
superstring with manifest spacetime supersymmetry
was introduced which uses pure spinors as worldsheet ghosts
\ref\superp{N. Berkovits, {\it Super-Poincar\'e Covariant Quantization of the
Superstring}, JHEP 04 (2000) 018, hep-th/0001035.}.
Since the worldsheet action is quadratic, it is straightforward to
compute manifestly super-Poincar\'e covariant
$N$-point tree amplitudes using this formalism and, last year, 
it was shown how to compute multiloop amplitudes \ref\loopold
{N. Berkovits, {\it Multiloop Amplitudes and Vanishing Theorems using
the Pure Spinor Formalism for the Superstring}, JHEP 0409 (2004) 047,
hep-th/0406055.}. In addition to proving various vanishing theorems
related to perturbative finiteness and S-duality \loopold, super-Poincar\'e
covariant massless four-point
one-loop \loopold\ and two-loop \ref\twoloop{N. Berkovits,
{\it Super-Poincar\'e Covariant Two-Loop Superstring Amplitudes},
hep-th/0503197.}\ amplitudes were explicitly computed.

To check consistency of the new formalism, it is useful to compare these
amplitudes with those amplitudes 
that have also been computed using the RNS and GS formalisms.
For massless $N$-point tree amplitudes involving
four or fewer Ramond states and an arbitrary number of Neveu-Schwarz
states, the equivalence
with the RNS computation was proven in \ref\vallilo{N. Berkovits and 
B.C. Vallilo, {\it Consistency of Super-Poincar\'e Covariant
Superstring Tree Amplitudes}, JHEP 07 (2000) 015, hep-th/0004171.}.
And for massless four-point one-loop amplitudes, the equivalence with
the RNS and GS computations was proven in \ref\aguel{L. Anguelova,
P.A. Grassi and P. Vanhove, {\it Covariant One-Loop Amplitudes in $D=11$},
Nucl. Phys. B702 (2004) 269, hep-th/0004171.}.

For massless four-point two-loop amplitudes, computations 
have only been performed using the RNS formalism 
for the case when all four
external states are Neveu-Schwarz 
\ref\phong{E. D'Hoker and D.H. Phong, {\it Two Loop Superstrings, 
1. Main Formulas}, Phys. Lett. B529 (2002)
241, hep-th/0110247\semi
E. D'Hoker and D.H. Phong,
{\it Two Loop Superstrings, 2. The Chiral Measure on Moduli Space}, Nucl.
Phys. B636 (2002) 3, hep-th/0110283\semi
E. D'Hoker and D.H. Phong,
{\it Two Loop Superstrings, 3. Slice Independence and Absence
of Ambiguities}, Nucl. Phys. B636 (2002) 61, hep-th/0111016\semi
E. D'Hoker and D.H. Phong,
{\it Two Loop Superstrings, 4. The Cosmological Constant and Modular
Forms},
Nucl. Phys. B639 (2002) 129, hep-th/0111040\semi
E. D'Hoker and D.H. Phong,
{\it Two Loop Superstrings, 5. Gauge-Slice Independendence of the N-Point
Function}, hep-th/0501196\semi
E. D'Hoker and D.H. Phong,
{\it Two Loop Superstrings, 6. Non-Renormalization Theorems and the
Four-Point Function}, hep-th/0501197.}
\ref\zhuold{
R. Iengo and C.-J. Zhu, {\it
Two Loop Computation of the Four-Particle
Amplitude in the Heterotic String}, Phys. Lett. B212 (1988) 313\semi
R. Iengo and C.-J. Zhu, {\it Explicit Modular Invariant Two-Loop Superstring
Amplitude Relevant for $R^4$}, JHEP 06 (1999) 011, hep-th/9905050
\semi R. Iengo, {\it
Computing the $R^4$ Term at Two Superstring Loops}, JHEP 0202 (2002) 035,
hep-th/0202058\semi
Z.-J. Zheng, J.-B. Wu and C.-J. Zhu, {\it
Two-Loop Superstrings in Hyperelliptic Language I: the Main
Results}, Phys. Lett. B559 (2003) 89, hep-th/0212191\semi
Z.-J. Zheng, J.-B. Wu and C.-J. Zhu, {\it
Two-Loop Superstrings in Hyperelliptic Language II: the Vanishing
of the Cosmological Constant and the Non-Renormalization Theorem},
Nucl. Phys. B663 (2003) 79, hep-th/0212198\semi
Z.-J. Zheng, J.-B. Wu and C.-J. Zhu, {\it
Two-Loop Superstrings in Hyperelliptic Language III: the Four-Particle
Amplitude}, Nucl. Phys. B663 (2003) 95, hep-th/0212219\semi
W.-J. Bao and C.-J. Zhu, {\it Comments on Two-Loop Four-Particle
Amplitude in Superstring Theory}, JHEP 0305 (2003) 056, hep-th/0303152.}.
Because of the need to sum over 
spin structures and include surface term contributions, these RNS
computations are extremely complicated. On the other hand, computation
of massless four-point two-loop amplitudes 
using the super-Poincar\'e covariant formalism is easy since
the fermionic worldsheet variables only contribute through their zero
modes\twoloop. The final result is quite simple and is expressed
as a superspace integral in terms of the ten-dimensional super-Yang-Mills
and supergravity superfields.

In this paper, the integral over superspace will be explicitly performed
for the case when all external states are in the Neveu-Schwarz sector.
The amplitude will then be shown to coincide with the RNS result of
\phong\zhuold. 

\newsec{Comparison of Two-Loop Amplitudes}

As derived in \twoloop\ using the methods of \loopold,
the four-point two-loop Type IIB amplitude 
computed using the pure spinor formalism is
\eqn\purespinor{
{\cal A}=\int d^2\Omega_{11}d^2\Omega_{12}d^2\Omega_{22}\prod_{R=1}^{4}
\int d^2z_R{{\exp{\left(-\Sigma_{R,S=1}^4k_R\cdot k_S
G(z_R,z_S)\right)} }\over{\left(\det Im\Omega\right)^5}} }
$$\Big|\left(\int
d^5\theta\right)^{\alpha\beta\gamma} 
(\gamma^{mnpqr})_{\alpha\beta}\gamma^{s}_{\gamma\delta}$$
$$\left(
{\cal F}^1_{mn}(\theta){\cal F}^2_{pq}(\theta){\cal F}^3_{rs}(\theta)
W^{4\delta}(\theta)
\Delta(z_1,z_3)\Delta(z_2,z_4)+\rm{perm}(1234)\right)\Big|^2 $$
where 
$\Omega_{CD}$ is the genus-two period matrix for $C,D=1$ to 2, 
$\Delta(y,z)= \epsilon^{CD}\omega_C(y)\omega_D(z)$,
$\omega_C$ are the two holomorphic one-forms,
$G(y,z)$ is the scalar Green's function, $|~~|^2$ denotes
the product of left and right-moving open superstring expressions, 
$W^{R\a}(\t)$ and 
${\cal F}_{mn}^R(\t)$
are the linearized
spinor and vector super-Yang-Mills superfield-strengths for the
$R^{th}$ external state with momentum $k^m_R$ satisfying $k_R\cdot k_R=0$, 
\eqn\thetaintegral{\left(\int d^5\theta\right)^{\alpha\beta\gamma}=
(T^{-1})^{\alpha\beta\gamma}_{\rho_1{\ldots}\rho_{11}}
\epsilon^{\rho_1{\ldots} \rho_{16}}
{\p\over{\p\theta^{\rho_{12}}}}{\ldots}{\p\over{\partial
\theta^{\rho_{16}}}},}
and $(T^{-1})^{\a\b\g}_{\rho_1 ...\rho_{11}}$ is
a Lorentz-invariant tensor which is antisymmetric
in $[\rho_1 ...\rho_{11}]$ and symmetric and $\g$-matrix traceless
in $(\a\b\g)$. Up to an overall normalization constant,
\eqn\deftinv{(T^{-1})^{\a\b\g}_{\rho_1 ...\rho_{11}} =
\e_{\rho_1 ...\rho_{16}}(\g^m)^{\k\rho_{12}}
(\g^n)^{\s\rho_{13}}
(\g^p)^{\tau\rho_{14}}
(\g_{mnp})^{\rho_{15}\rho_{16}} (\d^{(\a}_{\k}\d^\b_\s\d^{\g)}_{\tau}
-{1\over{40}}\g_q^{(\a\b}\d^{\g)}_\k \g^q_{\s\tau}).}

Comparing \purespinor\ with the RNS result of \phong\zhuold\ and ignoring the
Ramond component fields in the superfields $W^{R\a}$ and
${\cal F}_{mn}^R$, one finds that the
results coincide if 
\eqn\coincide{ 
t_8^{m_1 n_1 ... m_4 n_4} F^1_{m_1 n_1} F^2_{m_2 n_2} F^3_{m_3 n_3} 
F^4_{m_4 n_4} 
 {\cal Y} =
\left(\int
d^5\theta\right)^{\alpha\beta\gamma} 
(\gamma^{mnpqr})_{\alpha\beta}\gamma^{s}_{\gamma\delta}}
$$\left(
{\cal F}^1_{mn}(\theta){\cal F}^2_{pq}(\theta){\cal F}^3_{rs}(\theta)
W^{4\delta}(\theta)
\Delta(z_1,z_3)\Delta(z_2,z_4)+\rm{perm}(1234)\right), $$
where $t_8 F^1 F^2 F^3 F^4$ 
is the well-known kinematic factor 
appearing also in four-point tree-level and one-loop computations, 
$F_{mn}^R$ is the ordinary linearized Yang-Mills field-strength of the $R^{th}$
external state, and 
\eqn\defY{{\cal Y}= (k_1-k_2)\cdot (k_3-k_4)\Delta(z_1,z_2)\Delta(z_3,z_4)}
$$ +(k_1-k_3)\cdot (k_2-k_4)\Delta(z_1,z_3)\Delta(z_2,z_4)
	+(k_1-k_4)\cdot (k_2-k_3)\Delta(z_1,z_4)\Delta(z_2,z_3).$$

To evaluate the right-hand side of \coincide, it is convenient
to use the notation
\eqn\notT{
(T^{-1})^{\a\b\g}_{\rho_1{\ldots}\rho_{11}}\e^{\rho_1{\ldots} \rho_{16}}
 \longrightarrow 
\langle \lambda^{\alpha}\lambda^{\beta}\lambda^{\gamma}
\theta^{\rho_{12}}\theta^{\rho_{13}}
\theta^{\rho_{14}}\theta^{\rho_{15}}\theta^{\rho_{16}}\rangle}
where $\l^\a$ is a pure spinor,
which is motivated by the original definition of 
$(T^{-1})^{\a\b\g}_{\rho_1{\ldots}\rho_{11}}$
in the amplitude computations of \superp.
Using that ${{\p}\over{\p\t^\a}}$ can be substituted by
$D_\a = 
{\p\over{\p\t^\a}} + \half (\g^m\t)_\a \p_m$ because of conservation
of momentum, the right-hand side of \coincide\ can be written as
\eqn\ddddd{
\Delta(z_1,z_3)\Delta(z_2,z_4)
\langle(\lambda\gamma^{mnpqr}\lambda)(\lambda\gamma^s)_{\delta}
(\t^5)^{[\rho_1 ... \rho_5]}\rangle ~ (D^5)_{[\rho_1 ...\rho_5]}
\left(
{\cal F}_{mn}^1{\cal F}^2_{pq}{\cal F}^3_{rs}W^{4\delta}\right)}
$$+\quad{\rm permutations }\quad {\rm of }\quad (1234).$$
Note that throughout this paper, we will use the antisymmetrization
convention that
\eqn\conv{
f_{[a_1 ... a_N]} = {{\sum_{{\rm perm}(1...N)} (-1)^{sign(\s)}
f_{a_{\s(1)} ... a_{\s(N)}}  }\over {N!}}.}

Since we only want to consider the Neveu-Schwarz sector and 
${\cal F}_{mn}$ is bosonic while $W^\a$ is
fermionic, the only contribution to this 
computation comes 
from terms in which an even number of $D$'s act upon each ${\cal F}$ and an
odd number of $D$'s act on $W$. 
One therefore has
\eqn\fived{D^5({\cal F}^1_{mn}{\cal F}^2_{pq} {\cal F}^3_{rs}W^{4\d})=
{\cal F}^1_{mn}{\cal F}^2_{pq}{\cal F}^3_{rs} ~D^5 W^{4\d} + }
$$5
[(D^4{\cal F}^1_{mn}){\cal F}^2_{pq}{\cal F}^3_{rs}
+{\cal F}^1_{mn}(D^4{\cal F}^2_{pq}){\cal F}^3_{rs}
+{\cal F}^1_{mn}{\cal F}^2_{pq}(D^4{\cal F}^3_{rs})]~DW^{4\d}$$
$$+10[
(D^2{\cal F}^1_{mn}){\cal F}^2_{pq}{\cal F}^3_{rs}+
{\cal F}^1_{mn}(D^2{\cal F}^2_{pq}){\cal F}^3_{rs} +
{\cal F}^1_{mn}{\cal F}^2_{pq}(D^2{\cal F}^3_{rs})]~D^3W^{4\d}$$
$$+30[(D^2{\cal F}^1_{mn})(D^2{\cal F}^2_{pq}){\cal F}^3_{rs}+
(D^2{\cal F}^1_{mn}){\cal F}^2_{pq}(D^2{\cal F}^3_{rs})
+{\cal F}^1_{mn}(D^2{\cal F}^2_{pq})(D^2{\cal F}^3_{rs})]~DW^{4\d},$$
where the spinor indices on the five $D$'s are antisymmetrized and
the combinatoric factors in \fived\ come
from the different ways of splitting up these five indices.

After using $D_{\a}W^{\d}={1\over 4}(\g^{tu})_{\a}^{~\d}{\cal F}_{tu}$,
\ddddd\ is proportional to
\eqn\fourd{
\Delta(z_1,z_3)\Delta(z_2,z_4)
\langle (\lambda\gamma^{mnpq[r}\lambda)(\lambda\gamma^{s]}\g^{tu}\theta)(\t)^4
\rangle  ~{\Big[}
{\cal F}^1_{mn}{\cal F}^2_{pq}{\cal F}^3_{rs} ~D^4 {\cal F}^4_{tu} + }
$$5
[(D^4{\cal F}^1_{mn}){\cal F}^2_{pq}{\cal F}^3_{rs}
+{\cal F}^1_{mn}(D^4{\cal F}^2_{pq}){\cal F}^3_{rs}
+{\cal F}^1_{mn}{\cal F}^2_{pq}(D^4{\cal F}^3_{rs})]~{\cal F}^4_{tu}$$
$$+10[
(D^2{\cal F}^1_{mn}){\cal F}^2_{pq}{\cal F}^3_{rs}+
{\cal F}^1_{mn}(D^2{\cal F}^2_{pq}){\cal F}^3_{rs} +
{\cal F}^1_{mn}{\cal F}^2_{pq}(D^2{\cal F}^3_{rs})]~D^2 {\cal F}^4_{tu}$$
$$+30[(D^2{\cal F}^1_{mn})(D^2{\cal F}^2_{pq}){\cal F}^3_{rs}+
(D^2{\cal F}^1_{mn}){\cal F}^2_{pq}(D^2{\cal F}^3_{rs})
+{\cal F}^1_{mn}(D^2{\cal F}^2_{pq})(D^2{\cal F}^3_{rs})]~{\cal F}^4_{tu}
{\Big]}$$
$$+\quad{\rm permutations }\quad {\rm of }\quad (1234),$$
where the spinor indices on the four $D$'s are antisymmetrized and
contracted with the spinor indices on $(\t)^4$. 
As will be explained later, all terms in \fourd\
containing factors of $D^4 {\cal F}$
will not contribute to the amplitude.

Using the relations $D_\a {\cal F}^{mn} = 2k^{[m}\g^{n]}_{\a\b} W^\b$
and $D_\b W^\g = {1\over 4}(\g^{mn})_\b^{~\g} {\cal F}_{mn}$ where $k^m$ is the
momentum, one can express 
$D^2 {\cal F}_{mn}$
and $D^4 {\cal F}_{mn}$ in terms of ${\cal F}_{mn}$ as
\eqn\relation{
D_{\beta}D_{\alpha}{\cal F}_{mn}=
-{1\over 2}k_{[m}\left(\g_{n]}\g^{tu}\right)_{\a\beta}{\cal F}_{tu},}
$$
D_\d D_\g D_{\beta}D_{\alpha}{\cal F}_{mn}=
{1\over {4}}k_{[m}\left(\g_{n]}\g^{tu}\right)_{\a\b}
k_{t}\left(\g_{u}\g^{vw}\right)_{\g\d}
{\cal F}_{vw}.$$

Plugging \relation\ into \fourd\ and replacing ${\cal F}_{mn}^R$ with its
$\theta=0$ component $F_{mn}^R$, one obtains that the right-hand side
of \coincide\ is proportional to
\eqn\nextf{\Delta(z_1,z_3)\Delta(z_2,z_4)
\langle (\lambda\gamma^{mnpq[r}\lambda)(\lambda\gamma^{s]}\g^{tu}\theta)
(\t\g^{fgh}\t)(\t^{jkl}\t)
\rangle }
$$ {\Big[}
k^4_t k^4_g \eta_{hj}\eta_{uf} F^1_{mn} F^2_{pq} F^3_{rs} F^4_{kl} + $$
$$5 \eta_{hj}
[k^1_m k^1_g \eta_{nf} F^1_{kl} F^2_{pq} F^3_{rs}
+ k^2_p k^2_g \eta_{qf} F^1_{mn} F^2_{kl} F^3_{rs}
+ k^3_r k^3_g\eta_{sf} F^1_{mn} F^2_{pq} F^3_{kl}]~ F^4_{tu}$$
$$+10 k^4_t \eta_{uj}[
k^1_m \eta_{nf} F^1_{gh} F^2_{pq} F^3_{rs}+
k^2_p \eta_{qf} F^1_{mn} F^2_{gh} F^3_{rs} +
k^3_r \eta_{sf} F^1_{mn} F^2_{pq} F^3_{gh}]~ F^4_{kl}$$
$$+30[k^1_m k^2_p \eta_{nf}\eta_{qj} F^1_{gh} F^2_{kl} F^3_{rs}+
k^1_m k^3_r \eta_{nf}\eta_{sj} F^1_{gh} F^2_{pq} F^3_{kl} $$
$$
+ k^2_p k^3_r \eta_{qf}\eta_{sj} F^1_{mn} F^2_{gh} F^3_{kl}]~ F^4_{tu}
{\Big]}$$
$$+\quad{\rm permutations }\quad {\rm of }\quad (1234).$$

To check if \nextf\ reproduces the desired $t_8 F^1 F^2 F^3 F^4$
contractions, one needs to evaluate
\eqn\eval{\langle (\lambda\gamma^{mnpqr}\lambda)(\lambda\gamma^s\g^{tu}\theta)
(\t\g^{fgh}\t)(\t^{jkl}\t)\rangle =}
$$\langle (\lambda\gamma^{mnpqr}\lambda)(\lambda\gamma^{stu}\theta)
(\t\g^{fgh}\t)(\t^{jkl}\t)\rangle +
2\langle (\lambda\gamma^{mnpqr}\lambda)\eta^{s[t}(\lambda\gamma^{u]}\theta) 
(\t\g^{fgh}\t)(\t^{jkl}\t)\rangle .$$
Fortunately, the properties of pure spinors and the symmetries of \eval\
make this a straightforward task. Since \eval\ contains fourteen vector
indices and is Lorentz invariant,
it can be expressed in terms of linear combinations of products
of seven $\eta_{pq}$ tensors, or products of one ten-dimensional
$\epsilon$ tensor
and two $\eta_{pq}$ tensors. However, since the four-point amplitude
only involves three independent momenta and four polarizations, the
ten-dimensional $\e$ tensor cannot contribute to the four-point amplitude.
One can easily check that the only
possible linear combination of $\eta_{pq}$ tensors which has
the appropriate symmetries is
\eqn\iind{\langle(\l\g^{mnpqr}\l)(\l\g^{stu}\t)(\t\g_{fgh}\t)
(\t\g_{jkl}\t)\rangle=}
$$
X\Big[\d^{[s}_{[f}\d^t_g \eta^{u][m}\d^n_{h]}\d^p_{[j}\d^q_k\d^{r]}_{l]}
 +\d^{[s}_{[j}\d^t_k \eta^{u][m}\d^n_{l]}\d^p_{[f}\d^q_g\d^{r]}_{h]} $$
$$
-A\eta^{v[s}\d^t_{[f}\eta^{u][m}\d_g^n\eta_{h][j}\d^p_k
\d^q_{l]}\d_v^{r]}-A
\eta^{v[s}\d^t_{[j}\eta^{u][m}\d_k^n\eta_{l][f}\d^p_g
\d^q_{h]}\d_v^{r]} \Big],$$
\eqn\ind{\langle(\l\g^{mnpqr}\l)(\l\g^{u}\t)(\t\g_{fgh}\t)
(\t\g_{jkl}\t)\rangle = 
Z{\Big[}\d^{[m}_{[j} \d^n_k \d^p_{l]}\d^q_{[f}\d^{r]}_g \d^u_{h]}
+\d^{[m}_{[f} \d^n_g \d^p_{h]}\d^q_{[j}\d^{r]}_k \d^u_{l]}}
$$
-B\d^{[m}_{[j}\d^n_k \eta_{l][f}\d^p_g\d^q_{h]}\eta^{r]u}
-B\d^{[m}_{[f}\d^n_g \eta_{h][j}\d^p_k\d^q_{l]}\eta^{r]u}
\Big],$$
where $A$, $B$, $X$ and $Z$ are constants. The
coefficients $A$ and $B$ 
are determined from the pure spinor conditions 
\eqn\requirem{
\eta_{ms}\eta_{nt}(\l\g^{mnpqr}\l)(\l\g^{stu}\t)(\t\g_{fgh}\t)(\t\g_{jkl}\t)
=0,}
\eqn\requiren{
\eta_{mu}(\l\g^{mnpqr}\l)(\l\g^{u}\t)(\t\g_{fgh}\t)(\t\g_{jkl}\t)=0}
to be $A=1$ and $B={1\over 2}$.
And the constants $X$ and $Z$ are determined to be
$X= 3Z = -{12\over{35}}$
from
the relation
$$(\l\g^{mnpqr}\l)(\t\g_{npq}\t)=96(\l\g^m\t)(\l\g^r\t)$$
and the normalization condition that  
$$\langle(\l\g^m\t)(\l\g^n\t)(\l\g^p\t)(\t\g_{mnp}\t)\rangle =1.$$
Note that \iind\ and \ind\ imply that
\eqn\explain{
\langle(\l\g^{mnpqr}\l)(\l\g^{stu}\t)(\t\g_{fgh}\t)(\t\g_{jkl}\t)\rangle
\eta^{hj}=
\langle(
\l\g^{mnpqr}\l)(\l\g^{u}\t)(\t\g_{fgh}\t)(\t\g_{jkl}\t)\rangle\eta^{hj}=0,}
so there is no contribution from the
second and third lines of \nextf\ which come from
terms in \fourd\ with a $D^4{\cal F}$ factor.

Using the above formul{\ae}, it is straightforward to evaluate \nextf\
with the help of the mathematica package GAMMA \ref\GAM{U. Gran,
{\it GAMMA: A Mathematica Package for Performing Gamma-Matrix Algebra
and Fierz Transformations in Arbitrary Dimensions}, hep-th/0105086.}
for performing the tedious sum over the antisymmetrized deltas.\foot
{We are 
very greatful to Dr. Ulf Gran, the author of
the GAMMA package, for providing by request an efficient function to expand the
antisymmetrized deltas, 
which is not contained in the version available to
download at http://www.mth.kcl.ac.uk/\~{}ugran/.}
Writing $F_{mn}^R = k_m^R e_n^R - k_n^R e_m^R$ where $e_m^R$ is 
the polarization tensor satisfying $\eta^{mn} k_m^R e_n^R =0$,
and summing over all permutations of the $(1234)$ indices, 
one obtains an expression containing approximately 250 terms.
Using momentum conservation and expressing contractions of momenta 
in terms of the Mandelstam variables 
$s= -2(k^1\cdot k^2)$,
$t= -2(k^2\cdot k^3)$ and
$u= -2(k^1\cdot k^3)$, 
one obtains that the right-hand side of \coincide\
is proportional to
$\Delta(z_1,z_2)\Delta(z_3,z_4)$ multiplied by
\eqn\firstterm{ + 2(k^2\cdot e^3)(k^2\cdot e^4)(e^1\cdot e^2)t^2 
   + 2(k^2\cdot e^4)(k^4\cdot e^3)(e^1\cdot e^2)t^2 }
$$ - 2(k^2\cdot e^4)(k^3\cdot e^2)(e^1\cdot e^3)t^2
   + 2(k^3\cdot e^4)(k^4\cdot e^2)(e^1\cdot e^3)t^2 $$ 
$$ - 2(k^2\cdot e^3)(k^4\cdot e^2)(e^1\cdot e^4)t^2 
   - 2(k^4\cdot e^2)(k^4\cdot e^3)(e^1\cdot e^4)t^2 $$
$$ + 2(k^2\cdot e^4)(k^3\cdot e^1)(e^2\cdot e^3)t^2 
   + 2(k^2\cdot e^3)(k^4\cdot e^1)(e^2\cdot e^4)t^2 $$
$$ + 2(k^3\cdot e^1)(k^4\cdot e^3)(e^2\cdot e^4)t^2 
   + 2(k^4\cdot e^1)(k^4\cdot e^3)(e^2\cdot e^4)t^2 $$
$$ - 2(k^3\cdot e^1)(k^4\cdot e^2)(e^3\cdot e^4)t^2 
   + 2(k^2\cdot e^3)(k^3\cdot e^4)(e^1\cdot e^2)tu $$
$$ - 2(k^2\cdot e^4)(k^4\cdot e^3)(e^1\cdot e^2)tu 
   - 2(k^3\cdot e^2)(k^3\cdot e^4)(e^1\cdot e^3)tu $$
$$ - 2(k^3\cdot e^4)(k^4\cdot e^2)(e^1\cdot e^3)tu
   + 2(k^3\cdot e^2)(k^4\cdot e^3)(e^1\cdot e^4)tu $$
$$ + 2(k^4\cdot e^2)(k^4\cdot e^3)(e^1\cdot e^4)tu 
   + 2(k^3\cdot e^1)(k^3\cdot e^4)(e^2\cdot e^3)tu $$
$$ + 2(k^3\cdot e^4)(k^4\cdot e^1)(e^2\cdot e^3)tu
   - 2(k^3\cdot e^1)(k^4\cdot e^3)(e^2\cdot e^4)tu $$
$$ - 2(k^4\cdot e^1)(k^4\cdot e^3)(e^2\cdot e^4)tu 
   - 2(k^3\cdot e^2)(k^4\cdot e^1)(e^3\cdot e^4)tu $$
$$ + 2(k^3\cdot e^1)(k^4\cdot e^2)(e^3\cdot e^4)tu 
   - 2(k^2\cdot e^3)(k^2\cdot e^4)(e^1\cdot e^2)u^2 $$
$$ - 2(k^2\cdot e^3)(k^3\cdot e^4)(e^1\cdot e^2)u^2
   + 2(k^2\cdot e^4)(k^3\cdot e^2)(e^1\cdot e^3)u^2 $$
$$ + 2(k^3\cdot e^2)(k^3\cdot e^4)(e^1\cdot e^3)u^2 
   + 2(k^2\cdot e^3)(k^4\cdot e^2)(e^1\cdot e^4)u^2 $$
$$ - 2(k^3\cdot e^2)(k^4\cdot e^3)(e^1\cdot e^4)u^2
   - 2(k^2\cdot e^4)(k^3\cdot e^1)(e^2\cdot e^3)u^2 $$
$$ - 2(k^3\cdot e^1)(k^3\cdot e^4)(e^2\cdot e^3)u^2 
   - 2(k^3\cdot e^4)(k^4\cdot e^1)(e^2\cdot e^3)u^2 $$
$$ - 2(k^2\cdot e^3)(k^4\cdot e^1)(e^2\cdot e^4)u^2
   + 2(k^3\cdot e^2)(k^4\cdot e^1)(e^3\cdot e^4)u^2 $$
$$ + (e^1\cdot e^2)(e^3\cdot e^4)t^2u 
   - (e^1\cdot e^4)(e^2\cdot e^3)t^2u $$
$$ - (e^1\cdot e^3)(e^2\cdot e^4)t^3 
   + (e^1\cdot e^3)(e^2\cdot e^4)tu^2 $$
$$ - (e^1\cdot e^2)(e^3\cdot e^4)tu^2
   + (e^1\cdot e^4)(e^2\cdot e^3)u^3, $$
plus a second term multiplying $\Delta(z_1,z_3)\Delta(z_2,z_4)$ 
which is obtained from \firstterm\ by switching $2$ with $3$ 
and $s$ with $u$,
plus a third term multiplying
$\Delta(z_1,z_4) \Delta(z_3,z_2)$ which is obtained from
\firstterm\ by switching $2$ with $4$ and $s$ with $t$.
Expanding $t_8 F^1 F^2 F^3 F^4$ in terms of polarizations
and momenta, one can check that each of these three terms 
is proportional to
$ (t_8 F^1 F^2 F^3 F^4)$, and that the sum of the terms is
equal to
$(t_8 F^1 F^2 F^3 F^4)$ multiplied by
$$c[(t-u)\Delta(z_1,z_2) \Delta(z_3,z_4)
+ (t-s)\Delta(z_1,z_3) \Delta(z_2,z_4)
+ (s-u)\Delta(z_1,z_4) \Delta(z_3,z_2)]$$
where $c$ is a constant factor.
So it has been proven that the four-point two-loop amplitude
computed in \twoloop\ coincides with the RNS result of \phong\zhuold.

\vskip 15pt
{\bf Acknowledgements:} NB would like to thank  
CNPq grant 300256/94-9, 
Pronex 66.2002/1998-9,
and FAPESP grant 04/11426-0
for partial financial support, and CRM would like to thank
Dr.~Ulf Gran for his assistance with the GAMMA package, and
FAPESP grant 04/13290-8 for partial financial support. 

\listrefs

\end